\documentclass[conference]{IEEEtran}
\IEEEoverridecommandlockouts

\usepackage{url}
\usepackage{array}
\usepackage{cancel}
\usepackage{siunitx}
\usepackage{graphicx}
\usepackage{textcomp}
\usepackage{multirow}
\usepackage{stfloats}
\usepackage{listings}
\usepackage{colortbl}
\usepackage[T1]{fontenc}
\usepackage[10pt]{moresize}
\usepackage[utf8]{inputenc}
\usepackage[noadjust]{cite}
\usepackage{amsmath,bm,amsfonts}
\usepackage{mathtools, cuted, nccmath}
\usepackage{algorithmicx, algpseudocode, algorithm}

\graphicspath{{./figs/}}

\def\BibTeX{{\rm B\kern-.05em{\sc i\kern-.025em b}\kern-.08em
		T\kern-.1667em\lower.7ex\hbox{E}\kern-.125emX}
}

\begin{document}
	\title{Hardware-Efficient Compound IC Protection with Lightweight Cryptography \\
		\thanks{This work was co-funded by the European Union and Estonian Research Council via project TEM-TA138, the Estonian Research Council Grant PRG2531, and Estonian Ministry of Defence under grant No 2-2/24/541-1.}}
	
	\author{
		\IEEEauthorblockN{Levent~Aksoy\IEEEauthorrefmark{2}, Muhammad~Sohaib~Munir\IEEEauthorrefmark{2} and Sedat~Akleylek\IEEEauthorrefmark{3}}
		\IEEEauthorblockA{\IEEEauthorrefmark{2}Department of Computer Systems, Tallinn University of Technology, Tallinn, Estonia\\
			\IEEEauthorrefmark{3}Institute of Computer Science, University of Tartu, Tartu, Estonia\\
			Email: levent.aksoy@taltech.ee, muhammad.munir@taltech.ee, sedat.akleylek@ut.ee}
	}
	
	\maketitle
	
	\begin{abstract}
		Over the years, many techniques have been introduced to protect integrated circuits (ICs) from hardware security threats that emerged in the globalized IC manufacturing supply chain, such as overproduction and piracy. However, most of these techniques have been rendered inefficient since they do not rely on provably secure algorithms. Moreover, the previously proposed techniques using cryptography algorithms lead to a significant increase in hardware complexity and are vulnerable to the removal and power analysis attacks. In this paper, we propose a compound IC protection mechanism that uses a lightweight cryptography algorithm with prominent logic locking and hardware obfuscation techniques. Experimental results show that the secure designs generated by the developed tool have significantly less hardware complexity when compared to those generated by previously proposed techniques using cryptography algorithms and are resilient to existing removal, algebraic, and logic locking attacks.
	\end{abstract}
	
	\begin{IEEEkeywords}
		hardware security, logic locking, lightweight cryptography, removal, algebraic, and logic locking attacks.
	\end{IEEEkeywords}
	
	\section{Introduction}

The increasing complexity of integrated circuits (ICs), aggressive \mbox{time-to-market} demands, and the high cost of building a foundry at advanced technology nodes have been pushing the IC design industry to the globalized IC design flow. However, such a horizontal model, which consists of many entities, paves the way for serious security threats, such as overproduction and piracy. Although many efficient techniques, including logic locking and hardware obfuscation, have been introduced to mitigate these threats~\cite{kamali22}, in this \mbox{cat-and-mouse} game, many of them have been broken because they do not rely on provably secure algorithms.

Cryptography algorithms are based on problems that cannot be solved in polynomial time on classical computers~\cite{paar10}. Since the conventional cryptosystems, such as RSA~\cite{rsa78}, ECC~\cite{koblitz87}, and AES~\cite{aes}, cannot meet the power, performance, and area requirements of internet of things (IoT) devices, including radio-frequency identification tags and sensors, many lightweight cryptography (LWC) algorithms have been introduced~\cite{thakor21}. The LWC algorithms use lightweight hardware operations and have been proven to be secure against cryptanalysis attacks~\cite{biryukov17}. 

For the protection of ICs from overproduction and piracy, the use of cryptosystems has been proposed in~\cite{yasin16_tcad, massad22, saha20}. However, the techniques of~\cite{yasin16_tcad, massad22} increase the design complexity significantly due to the selected cryptography algorithm and locking mechanism, as shown in Section~\ref{sec:results}. Moreover, while the technique of~\cite{yasin16_tcad} is broken by the proposed removal attack introduced in Section~\ref{subsec:pa}, the technique of~\cite{saha20} is already broken by a power analysis attack as shown in~\cite{roy24}. Thus, the \textit{research problem} is defined as finding a defense mechanism that can mitigate overproduction and piracy with a small area overhead while being resilient to all existing and possible attacks. To address this problem, we introduce a compound IC protection technique that uses an LWC algorithm in collaboration with prominent logic locking and hardware obfuscation techniques. It initially locks an original design using a prominent logic locking technique and an LWC algorithm, and then uses a hardware obfuscation technique to hide the connections between the LWC algorithm and the locked design. It can incorporate LWC algorithms, such as a version of ASCON-AEAD128~\cite{nist_ascon}, PRESENT~\cite{bogdanov07}, and SIMON~\cite{beaulieu13}, with the logic locking technique TTLock, also known as SFLL-HD$^0$~\cite{yasin17}, and the look-up table \mbox{(LUT)-based} obfuscation technique~\cite{baumgarten10}. It is implemented in a computer-aided design (CAD) tool called SOHNI. We also introduce specific attacks to break the proposed technique and discuss the computational difficulties of existing and possible attacks. Thus, the main \textit{contributions} of this paper are given as follows:
\begin{itemize}
	\item An IC protection technique against overproduction and piracy using an LWC algorithm along with prominent logic locking and hardware obfuscation techniques;
	\item An open-source CAD tool that automates the process of generating secure ICs;
	\item New attacks, including the removal attack that can extract the original design from the circuit locked by the previously proposed technique of~\cite{yasin16_tcad} and the algebraic attack that aims to find the secret key.
\end{itemize}

Experimental results show that the previously proposed techniques using cryptography algorithms~\cite{yasin16_tcad} and~\cite{massad22} lead to locked circuits with significantly larger area, reaching up to $11.6\times$ and $59.8\times$ on a large ITC'99 circuit~\cite{corno00}, respectively, when compared to those generated by SOHNI. It is also shown that the secure designs generated by SOHNI are resilient to the proposed removal and algebraic attacks, and \mbox{well-known} satisfiability (SAT)-based logic locking attack and its variants~\cite{subramanyan15, shamsi17, shen17}, and structural analysis attacks~\cite{aksoy24, limaye22}. 

The rest of this paper is organized as follows: Section~\ref{sec:background} gives the background concepts and related work. The proposed attacks and compound IC protection mechanism are described in Section~\ref{sec:technique}, and experimental results are presented in Section~\ref{sec:results}. Finally, Section~\ref{sec:conclusions} concludes the paper.

	\section{Background}
\label{sec:background}

This section introduces the background concepts on logic locking and block ciphers, and presents the related work on the use of cryptography algorithms in hardware security.

\begin{figure*}[t]
	\centerline{\includegraphics[width=18.0cm]{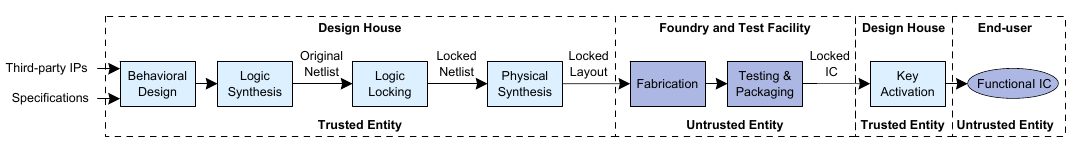}}
	\vspace*{-4mm}
	\caption{Conventional logic locking in the IC design flow.}
	\label{fig:iclock}
	\vspace*{-6mm}
\end{figure*}

\subsection{Logic Locking}
\label{subsec:ll}

Logic locking inserts an additional logic with key inputs into an original design such that the locked design behaves as the original one only when the secret key is provided. Otherwise, it generates a wrong output. It is generally applied at the \mbox{gate-level} in the IC design flow. As shown in Fig.~\ref{fig:iclock}, after the original design is locked and the layout of the locked design is generated, it is sent to the foundry without revealing the secret key. After the locked IC is fabricated by the foundry and delivered to the design house, values of the secret key are stored in a \mbox{tamper-proof} memory. Finally, the functional IC is sent to the market. 

\begin{figure}[t]
	\centerline{\includegraphics[width=7.0cm]{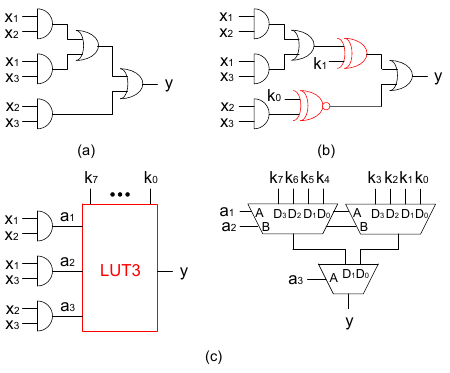}}
	\vspace*{-4mm}
	\caption{Logic locking on the majority circuit: (a)~original design; (b)~locked design using {\sc xor}/{\sc xnor} gates where the secret key is $k_1k_0 = 01$; (c)~locked design using LUTs where the secret key is $k_7\ldots k_1k_0 = 11111110$ and the design of LUT3 using multiplexors.}
	\label{fig:majority}
	\vspace*{-6mm}
\end{figure}

In logic locking, there exist two main threat models: i)~oracle-less (OL) and ii)~oracle-guided (OG). In the OL threat model, the adversary has only the locked netlist, which can be reverse-engineered from the layout at the foundry or from the functional IC obtained at the market. In the OG threat model, the adversary has the locked netlist and functional IC, which can be used as an oracle to apply inputs and observe outputs.

In the scenario for overproduction~\cite{xie17}, it is assumed that the adversary at an untrusted foundry can obtain the locked netlist by \mbox{reverse-engineering} the locked layout and can overproduce locked ICs. If the secret key of the locked design is discovered, the adversary can sell the unauthorized locked ICs. Otherwise, the overproduction of unauthorized locked ICs is futile. The adversary can use sophisticated attacks to find the secret key under the OL and OG threat models.

In the scenario for piracy~\cite{roy08}, it is assumed that the adversary at an untrusted foundry or an end-user can obtain the locked netlist by \mbox{reverse-engineering} the layout or functional IC, respectively. If the original design is extracted from the locked circuit, the adversary can use it for his/her benefit. To find the original design, the adversary either uses removal and reverse-engineering attacks that can identify the original design in the locked netlist or initially finds the secret key and then obtains the original design by embedding the values of the secret key into the locked design. 

\begin{figure}[t]
	\centerline{\includegraphics[width=8.5cm]{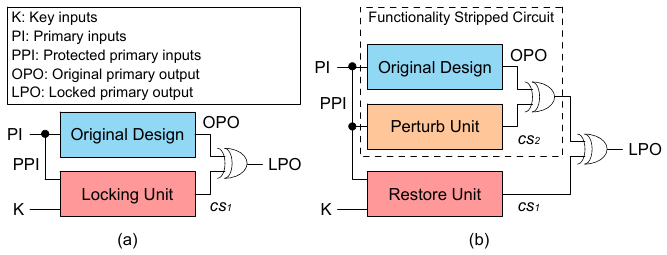}}
	\vspace*{-4mm}
	\caption{SAT-based attack resilient techniques: (a)~SFLT; (b)~DFLT.}
	\label{fig:sdflt}
	\vspace*{-6mm}
\end{figure}

Logic locking has been a prominent solution to mitigate overproduction and piracy, and a wide range of logic locking techniques and attacks have been proposed~\cite{kamali22}. The early techniques focus on hardware complexity and output corruption and use various logic gates with key inputs to lock an original design. Fig.~\ref{fig:majority} illustrates the logic locking using {\sc xor}/{\sc xnor} gates and LUTs, where \textit{LUT3} denotes the \mbox{3-input} LUT that can realize any logic function with three inputs. Note that the technique using LUTs not only locks a design using key inputs, but also obfuscates the original gates, e.g., a 3-input {\sc or} gate with the inputs of $a_1$, $a_2$, and $a_3$ in Fig.~\ref{fig:majority}(c). However, these techniques were broken by the \mbox{SAT-based} attack~\cite{subramanyan15} that iteratively eliminates wrong key(s) by finding distinguishing input patterns. To increase the number of iterations in the SAT-based attack, single flip locking techniques (SFLTs), e.g., Anti-SAT~\cite{xie19} and SARLock~\cite{yasin16}, and double flip locking techniques (DFLTs), e.g., TTLock~\cite{yasin17} and CAC~\cite{kaveh19}, force the \mbox{SAT-based} attack to eliminate only one wrong key in each iteration. As illustrated in Fig.~\ref{fig:sdflt}, while SFLTs use a single critical signal $cs_1$, which corrupts the original circuit for wrong keys, DFLTs use a critical signal $cs_2$ to corrupt the original design for a specific input pattern and use a critical signal $cs_1$ to correct this corruption. However, they were broken by structural analysis techniques~\cite{sirone19, zhaokun21, patnaik22, limaye22, aksoy24}, which identify each protected primary input associated with a key input and explore all traces of the protected input pattern, i.e., the secret key, set by the logic synthesis tool in the locking unit of an SFLT and in the perturb unit of a DFLT. 

\subsection{Block Ciphers}
\label{subsec:lwca}

A block cipher operates on a fixed-length bits called block. Although there is no reduction from the encryption process of a block cipher to any well-known NP-hard problems, it is hypothesized that the provable construction of its one-way function is impractical due to the employed confusion and diffusion techniques~\cite{paar10}. 

AES~\cite{aes} is a 128-bit block cipher based on the substitution-permutation network (SPN). It supports three key lengths, i.e., 128, 192, and 256, including 10, 12, and 14 rounds, respectively. It generates round keys that are added to the internal data in each round. It introduces confusion to ensure that changes in the individual internal bits propagate quickly across the data path using an \mbox{8-bit} to \mbox{8-bit} LUT with special mathematical properties, called \mbox{S-box}. It introduces diffusion by permuting and transforming the internal data. 

\begin{figure}[t]
	\centerline{\includegraphics[width=9.5cm]{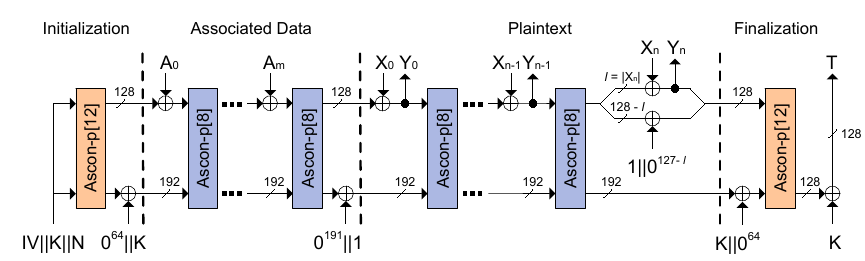}}
	\vspace*{-2mm}
	\caption{Encryption in ASCON-AEAD128.}
	\label{fig:ascon}
	\vspace*{-6mm}
\end{figure}

The authenticated encryption with associated data (AEAD), i.e., ASCON-AEAD128~\cite{nist_ascon}, takes a 128-bit key $K$, a \mbox{128-bit} nonce $N$, \mbox{variable-length} associated data $A$, and \mbox{variable-length} plaintext $X$ as inputs and generates the ciphertext $Y$, where the size of $X$ is equal to that of $Y$, and a 128-bit authentication tag $T$. Fig.~\ref{fig:ascon} outlines its four main phases during the encryption, i.e., initialization, associated data processing, plaintext processing, and finalization, where \mbox{\textit{Ascon-p[r]}} denotes its permutation block with the number of rounds $r \in \{8,12\}$ and $IV$ is the initialization vector. The permutation block is based on SPN and includes the constant addition, substitution, and diffusion layers. While the constant addition layer involves {\sc xor} operations, the substitution layer includes a 5-bit S-box, and the diffusion layer uses {\sc xor} and cyclic shift operations. In this work, ASCON-AEAD128 is used as a block cipher to encrypt a 128-bit plaintext $X$ with a 128-bit key $K$ and generate a 128-bit ciphertext $Y$. To do so, the nonce $N$ is set to a constant value, the entire associated data processing and finalization phases, and permutation blocks in the plaintext processing phase are omitted. This version is called ASCON throughout the paper.

PRESENT~\cite{bogdanov07} is a \mbox{64-bit} block cipher based on SPN. It supports two key lengths, i.e., 80 and 128, and has a fixed 31 rounds. While it uses a \mbox{4-bit} S-box for data confusion, it uses permutations for data diffusion. 

SIMON~\cite{beaulieu13} is based on a Feistel network, which implements the substitution and permutation alternately. It uses lightweight operations, such as {\sc and}, {\sc xor}, and cyclic shift. 

Table~\ref{tab:ciphers} summarizes the parameters used in these block ciphers. Observe that SIMON supports a wide range of block sizes $bs$ and key lengths $kl$ with various numbers of rounds $r$. 

\begin{table}[t]
	\footnotesize
	\centering
	\caption{Block cipher parameters.} 
	\vspace*{-3mm}
	\begin{tabular}{|l||c|c|c|}
		\hline  
		\multirow{2}{*}{Cipher} & Block Size & Key Length & Number of Rounds \\
		& ($bs$)     & ($kl$)     & ($r$) \\
		\hline
		\hline
		\multirow{10}{*}{SIMON}  & 32                   & 64  & 32 \\
		\cline{2-4}
		& \multirow{2}{*}{48}  & 72  & \multirow{2}{*}{36} \\
		&                      & 96  & \\
		\cline{2-4}
		& \multirow{2}{*}{64}  & 96  & 42 \\
		&                      & 128 & 44 \\
		\cline{2-4}
		& \multirow{2}{*}{96}  & 96  & 52 \\
		&                      & 144 & 54 \\
		\cline{2-4}
		& \multirow{3}{*}{128} & 128 & 68 \\
		&                      & 192 & 69 \\
		&                      & 256 & 72 \\
		\hline
		\multirow{2}{*}{PRESENT} & 64 & 80  & 31 \\
		& 64 & 128 & 31 \\
		\hline
		ASCON                    & 128 & 128 & 12* \\ 
		\hline
		\multirow{3}{*}{AES}     & 128 & 128 & 10 \\
		& 128 & 192 & 12 \\
		& 128 & 256 & 14 \\
		\hline
		\multicolumn{4}{l}{{*Denotes the number of rounds in its permutation block}} \\
	\end{tabular}
	\label{tab:ciphers}
	\vspace*{-6mm}
\end{table}

\subsection{Related Work}

Cryptosystems are proposed to be used for IC activation and user authentication, especially in security-sensitive systems and intellectual properties in~\cite{huang08}. In this scenario, an asymmetric key cryptosystem, i.e., ECC~\cite{koblitz87}, is used to unlock the protected circuit using the technique of~\cite{roy08}. To do so, the signature generated by the user's private key, which includes the physical characteristics of the IC obtained by a physical unclonable function, is validated through an encryption process by the user and a decryption process by the activation control unit. A similar approach is applied for FPGAs using a symmetric key cryptosystem, i.e., AES, in~\cite{simpson06}. The FORTIS framework~\cite{guin16} uses cryptosystems to handle the message integrity, authentication, and confidentiality during the secure transfer of the secret key of the locked design from the designer to the IC.  Moreover, AES is used to thwart the SAT-based attack~\cite{subramanyan15} while locking with another technique in~\cite{yasin16_tcad}. The method of~\cite{massad22} locks the original design based on the structure of a DFLT using Trivium~\cite{canniere06} as a puncturable pseudorandom function. The original design is hidden within multiple rounds of the PRESENT block cipher by configuring its \mbox{S-box} and using additional logic between the \mbox{S-box} and permutation block in~\cite{saha20}. However, depending on the hardware complexity of the original design, this technique cannot embed the entire design into a block cipher, and more importantly, it is broken by reducing the search space of possible secret key combinations and using a power analysis attack in~\cite{roy24}.

	\section{Proposed Attacks and Technique}
\label{sec:technique}

This section initially presents the technique of~\cite{yasin16_tcad} and introduces a removal attack to find the original design and an algebraic attack to find the secret key. Then, it introduces the proposed compound IC protection mechanism and discusses its resiliency to the existing and possible attacks.

\subsection{Previously Proposed Technique}
\label{subsec:ppt}

The technique of~\cite{yasin16_tcad} locks an original design using AES as follows:

\begin{enumerate}
	\item \textit{Generating block cipher}: Describe the block cipher when all its rounds are unrolled. Synthesize this behavioral description and obtain its \mbox{gate-level} netlist. Find a \textit{pattern}\footnote{A pattern denotes the values of variables, e.g., for an $n$-bit variable $Z$, it is $\mathbf{z}_{n-1}\ldots \mathbf{z}_1\mathbf{z}_0 = 1\ldots 01$.} of its $bs$-bit plaintext input, i.e., $\mathbf{X}$, and a pattern of its $kl$-bit key input, i.e., $\mathbf{K}$, randomly and compute the pattern of its $bs$-bit ciphertext output, i.e., $\mathbf{Y}$, based on $\mathbf{X}$ and $\mathbf{K}$. Synthesize the gate-level netlist after setting each input in $X$ to its associated value in $\mathbf{X}$, removing the input $X$ from the netlist and propagating the assigned values. 
	\item \textit{Generating the locked design}: For each output $y_i$ of the block cipher, where $0 \leq i \leq bs-1$, select an output of a gate in the original design randomly and connect it to an {\sc xor} ({\sc xnor}) gate with $y_i$ if $\mathbf{y_i}$ obtained in Step 1 is 0 (1). Synthesize the locked design to obfuscate the {\sc xor/xnor} gates in the original design connected to the outputs of the block cipher.
\end{enumerate}

\begin{figure}[t]
	\centerline{\includegraphics[width=7.5cm]{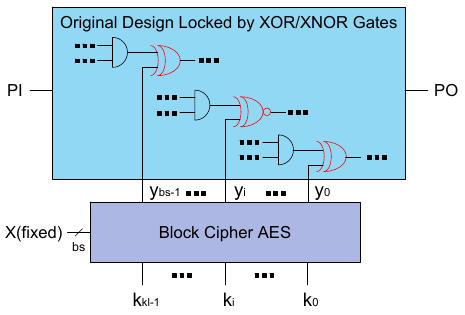}}
	\vspace*{-4mm}
	\caption{An original design locked by the technique of~\cite{yasin16_tcad}.}
	\label{fig:lllwc}
	\vspace*{-4mm}
\end{figure}

An original design locked by the technique of~\cite{yasin16_tcad} is illustrated in Fig.~\ref{fig:lllwc}, where $PI$ and $PO$ are the primary inputs and outputs of the original design, respectively. The number of {\sc xor/xnor} gates used to lock the original design is equal to the block size of AES, $bs$, the number of key bits of the locked design is equal to the key length of AES, $kl$, and the secret key is $\mathbf{K}$.

\subsection{Proposed Attacks}
\label{subsec:pa}

\begin{figure}[t]
	\centerline{\includegraphics[width=8.0cm]{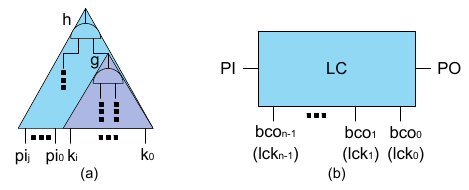}}
	\vspace*{-4mm}
	\caption{Details of the removal attack: (a)~Finding outputs of the block cipher; (b)~Generating the LC design.}
	\label{fig:ra}
	\vspace*{-6mm}
\end{figure}

It is claimed in~\cite{yasin16_tcad} that an adversary cannot find the original design since the logic synthesis obfuscates the connections between the block cipher and the original design. To refute this claim, we propose a \textbf{removal attack} described as follows: 
\begin{enumerate}
	\item \textit{Finding outputs of the cipher}: Determine the outputs of gates in the locked design given in Fig.~\ref{fig:lllwc} such that while all the inputs of the logic cone of a gate $g$ are key inputs in $K$, at least one of the inputs of the logic cone of the gate $h$, which is driven by $g$, is a primary input of the original design, as illustrated in Fig.~\ref{fig:ra}(a). Store the gate outputs like $g$ in a set called $BCO$, indicating the outputs of the block cipher. Assume that $|BCO|$ is equal to $n$.
	\item \textit{Removing the cipher}: Generate a design called $LC$ by removing the logic cones of each gate output in $BCO$ from the locked design. While its primary inputs are the primary inputs of the original design $PI$ and the gate outputs in $BCO$, its primary outputs are the primary outputs of the original design $PO$, as illustrated in Fig.~\ref{fig:ra}(b). 
	\item \textit{Running the SAT-based attack}: Replace the primary inputs of $LC$ denoting the gate outputs in $BCO$ with key inputs $LCK$, as illustrated in Fig.~\ref{fig:ra}(b). Run the SAT-based attack~\cite{subramanyan15} on $LC$ and oracle (functional IC) and determine the values of key inputs, i.e., $\mathbf{LCK}$.
	\item \textit{Finding the original design}: Synthesize $LC$ after setting each key input in $LCK$ to its associated value in $\mathbf{LCK}$, removing key input $LCK$ from $LC$ and propagating the assigned values. The resulting design is the original one. 
\end{enumerate}

Observe that the removal attack initially finds the block cipher and removes it from the locked design, keeping its outputs as key inputs. Then, it generates a problem of finding the values of key inputs of a design locked using {\sc xor/xnor} gates~\cite{roy08} and solves it using the \mbox{SAT-based} attack~\cite{subramanyan15}. Finally, it determines the original design by setting the key inputs to values found by the SAT-based attack, i.e., $\mathbf{LCK}$, and removing these key inputs using logic synthesis. 

To find the secret key of the design locked by the technique of~\cite{yasin16_tcad}, we propose an \textbf{algebraic attack} described as follows:
\begin{enumerate}
	\item \textit{Running the removal attack}: Apply the removal attack and determine $BCO$ and $\mathbf{LCK}$, which corresponds to the values of gate outputs in $BCO$, as given in Fig.~\ref{fig:ra}(b).
	\item \textit{Generating the block cipher}: Generate a design called $BC$ by adding the logic cone of each gate output in $BCO$ in the locked design. While its primary inputs are the key inputs of the block cipher $K$, its primary outputs are the gate outputs in $BCO$. 
	\item \textit{Running the SAT solver}: Describe $BC$ in a conjunctive normal form (CNF) formula, set values of its primary outputs to $\mathbf{LCK}$ as constraints, and run a SAT solver to find the values of key inputs that satisfy the constraints in the CNF formula. The resulting solution is the secret key of the block cipher, i.e., $\mathbf{K}$.
\end{enumerate}

Observe that the algebraic attack initially runs the removal attack and extracts the block cipher. Then, similar to~\cite{soos09}, it converts the problem of finding the secret key of a block cipher into a SAT problem when its input pattern $\mathbf{X}$ is already fixed and the output pattern $\mathbf{LCK}$ found by the removal attack is set as constraints. Note that it is only applicable when the block cipher is extracted from the locked design and its output pattern is determined.

\subsection{Proposed Compound IC Protection Technique}
\label{subsec:pt}

As shown in Section~\ref{sec:results}, the technique of~\cite{yasin16_tcad} described in Section~\ref{subsec:ppt} has two main drawbacks: i)~it leads to a significant increase in hardware complexity; ii)~it is vulnerable to the removal attack proposed in Section~\ref{subsec:pa}. To address the first drawback, an LWC algorithm can be used instead of AES. To address the second drawback, a DFLT can be used to lock the original design instead of the locking technique using {\sc xor/xnor} gates that is vulnerable to the SAT-based attack.  Moreover, a LUT-based obfuscation technique can be applied to hide the connections between the outputs of the block cipher and key inputs of a DFLT.

\begin{algorithm}[t]
	\small
	\caption{Proposed Compound IC Protection Technique}
	\begin{algorithmic}[1]
		\Statex \textbf{Inputs:} Original circuit $C$, LWC block cipher $BC$ with block size $bs$ and key length $kl$, and LUT size $m$.
		\Statex \textbf{Outputs:} Locked circuit $LC$ with the secret key \textbf{K}.
		\State ($\mathbf{X}, \mathbf{K}, \mathbf{Y}$) = \textit{Determine LWC IOs}($BC$, $bs$, $kl$) 
		\State $N_C$ = \textit{Logic Synthesis}($C$)
		\State $N_{DFLT}$ = \textit{Add DFLT}($N_C$, $bs$, $\mathbf{Y}$)
		\State $N_{LWC}$ = \textit{Add LWC Block Cipher}($N_{DFLT}$, $BC$, $\mathbf{X}$)
		\State ($N_{LUT}$, $\mathbf{K_{LUT}}$) = \textit{Add LUT-based Obfuscation}($N_{LWC}$, $m$)
		\State $LC$ = \textit{Logic Synthesis}($N_{LUT}$)
		\State $\mathbf{K}$ = $\mathbf{K} \cup \mathbf{K_{LUT}}$
	\end{algorithmic}
	\label{algo:ppt}
\end{algorithm}

\begin{figure}[t]
	\vspace*{-4mm}
	\centerline{\includegraphics[width=7.5cm]{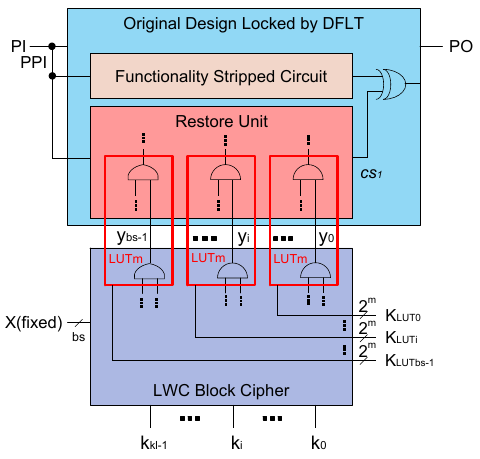}}
	\vspace*{-4mm}
	\caption{An original design locked by the proposed technique.}
	\label{fig:lllwc_lutobf}
	\vspace*{-6mm}
\end{figure}

Algorithm~\ref{algo:ppt} presents the steps of the proposed technique. It takes an original circuit $C$ described in either Verilog or bench format, the LWC block cipher $BC$ with block size $bs$ and key length $kl$, described in Verilog when all its rounds are unrolled, and the LUT size $m$ used in the LUT-based obfuscation as inputs. Initially, it randomly determines the values of the LWC block cipher inputs $X$ and $K$, i.e., $\mathbf{X}$ and $\mathbf{K}$, and computes the values of its output $Y$, i.e., $\mathbf{Y}$, based on $\mathbf{X}$ and $\mathbf{K}$. Then, the original circuit $C$ is synthesized and its gate-level netlist $N_C$ is obtained. A DFLT is applied to $N_C$, where a primary output, whose logic cone includes the number of primary inputs greater than or equal to $bs$, is chosen to be locked, its $bs$ primary inputs are selected to be the protected primary inputs, the protected input pattern is determined as $\mathbf{Y}$, and additional $bs$ key inputs are generated. The netlist locked by DFLT, i.e., $N_{DFLT}$, is extended by the LWC block cipher $BC$. In this case, $BC$ is synthesized when its input $X$ is fixed to $\mathbf{X}$ and its output $Y$ is connected to the $bs$ key inputs of $N_{DFLT}$. Finally, the LUT-based obfuscation is applied to the gates, which are used to connect the block cipher to the original design locked by DFLT through its restore unit. In this case, it is ensured that LUTs with a maximum size of $m$ with $2 \leq m \leq 6$ replace gates both in the block cipher and the restore unit of DFLT. The key inputs of LUTs, $K_{LUT_{i}}$ with $0 \leq i \leq bs-1$, are stored in $\mathbf{K_{LUT}}$. Finally, logic synthesis is applied to the netlist locked by DFLT, LWC block cipher, and LUT-based obfuscation, i.e., $N_{LUT}$, and the locked circuit $LC$ is obtained. The secret key is determined as the combination of the values of key inputs of the block cipher and LUTs. Fig.~\ref{fig:lllwc_lutobf} illustrates an original design locked by the proposed technique, where the red boxes denote LUTs. 

In the proposed technique, the LWC block cipher SIMON is preferred since it has various block sizes and key lengths with a minimum value of 32 and 64, respectively, as given in Table~\ref{tab:ciphers} and has the least hardware complexity among other block ciphers as shown in Section~\ref{sec:results}. TTLock is chosen as a DFLT since it is resilient to the SAT-based and removal attacks and its restore unit connected to the block cipher compares the protected primary inputs and the outputs of the block cipher. The maximum size of LUTs $m$ is preferred to be 4 to find a balance between security and hardware complexity. 

SOHNI automates the generation of secure designs using block ciphers ASCON, PRESENT, and SIMON, and is very flexible to include other LWC algorithms. It is equipped with the locking technique TTLock and the LUT-based obfuscation technique, and is very flexible to use other DFLTs and hardware obfuscation techniques. It utilizes Cadence Genus for logic synthesis and Cadence Conformal for equivalence checking between the original design and the locked design under the secret key. It is available at~\cite{sohni}.

\subsection{Discussion on Existing and Possible Attacks}
\label{subsec:dpa}

The proposed technique is resilient to the \mbox{SAT-based} attack and its variants due to TTLock, and also to the removal attack proposed in Section~\ref{subsec:pa} due to the TTLock and \mbox{LUT-based} obfuscation. Moreover, the brute-force and side-channel analysis attacks are inefficient due to a large number of key inputs and the full combinational unrolled structure of the LWC algorithm, as explained in~\cite{chawla19,moos20}, respectively. 

As a possible scenario, a structural analysis attack similar to those of~\cite{limaye22, aksoy24} can locate the critical signal $cs_1$ in a design locked by SOHNI given in Fig.~\ref{fig:lllwc_lutobf}. Also, it can extract the functionality stripped circuit and restore unit, and determine the protected primary inputs as the primary inputs of the restore unit other than the key inputs\footnote{The key inputs of a locked circuit are assumed to be known to the attacker since they are stored in a non-volatile memory.}. Furthermore, it can explore the traces of the protected input pattern of the protected primary inputs in the functionality stripped circuit, which corresponds to the block cipher output pattern $\mathbf{Y}$. However, first, such a structural analysis attack cannot determine the outputs of the block cipher since they are obfuscated by LUTs, and thus, the whole block cipher cannot be identified. Second, the secret key of the block cipher cannot be determined due to the provably secure LWC algorithm even if its output pattern $\mathbf{Y}$ is guessed from the traces of the protected input pattern.

Hence, these three techniques, i.e., DFLT, LWC block cipher, and LUT-based obfuscation, entwined together lead to a compound IC protection mechanism, which renders existing attacks unsuccessful in extracting the original design or in finding the secret key. 

	\section{Experimental Results}
\label{sec:results}

This section presents the gate-level synthesis results of ciphers and designs locked by the state-of-the-art logic locking techniques, SOHNI, and previously proposed techniques~\cite{yasin16_tcad, massad22}, and the results of attacks on these locked designs. 

\subsection{Ciphers}

The ciphers given in Section~\ref{subsec:lwca} are described at the register-transfer level (RTL) in Verilog when all their rounds are unrolled. Table~\ref{tab:lwc} presents the \mbox{gate-level} synthesis results of ciphers with similar block sizes and key lengths to make a fair comparison. In this table, \textit{area}, \textit{delay}, and \textit{power} are total area in $\mu m^2$, delay in the critical path in $ps$, and total power dissipation in $\mu W$, respectively, obtained after logic synthesis by Cadence Genus using a commercial 65\;nm gate library with the aim of area optimization. 

Observe from Table~\ref{tab:lwc} that while the AES design has the largest area, the ASCON design with the same block size and key length as AES, and the PRESENT design with the same key length as AES have significantly less hardware complexity than that of AES. In this case, the area of the AES design is $4.7\times$ and $8.9\times$ larger than that of ASCON and PRESENT, respectively. Various sizes of parameters in SIMON lead to designs with less hardware complexity than AES, ASCON,  and PRESENT. In this case, the area of AES, ASCON, and PRESENT when $kl$ is 80 is $25.5\times$, $5.4\times$, and $2.7\times$ larger than that of SIMON when $bs$ is 32, respectively. 

\begin{table}[t]
	\centering
	\caption{Synthesis results of unrolled block ciphers.}
	\vspace{-3mm}
	\footnotesize
	\begin{tabular}{|l||c|c|c||c|c|c|}
		\hline
		Cipher & $bs$ & $kl$ & $r$ & area & delay & power\\ 
		\hline \hline
		\multirow{2}{*}{SIMON}  & 32  & 64  & 32 & 9356  & 8673  & 2846  \\
		& 64  & 96  & 42 & 21368 & 9196  & 9722  \\
		\hline
		\multirow{2}{*}{PRESENT} & 64  & 80  & 31 & 25094  & 11204 & 10731 \\
		& 64  & 128 & 31 & 26809  & 11448 & 13535 \\
		\hline
		ASCON                    & 128 & 128 & 12* & 50400 & 7559 & 22544 \\
		\hline
		AES                      & 128 & 128 & 10 & 238691 & 26469 & 49413 \\
		\hline
		\multicolumn{7}{l}{{*Denotes the number of rounds in its permutation block}} \\
	\end{tabular}
	\label{tab:lwc}
	\vspace{-6mm}
\end{table}

\subsection{Defenses}

As the first experiment set, the combinational parts of four large circuits of the ITC'99 benchmark~\cite{corno00} are used to explore the impact of the state-of-the-art logic locking techniques, SOHNI, and previously proposed techniques~\cite{yasin16_tcad, massad22} on hardware complexity. 

These circuits are initially locked by the state-of-the-art logic locking techniques, i.e., using {\sc xor/xnor} gates, LUTs, Anti-SAT, and TTLock, with 128 key inputs using the CAD tool HIID~\cite{aksoy24_lats}. Table~\ref{tab:sota_ll} presents the gate-level synthesis results of original and locked circuits. 

Observe that while the logic locking technique using {\sc xor/xnor} gates leads to locked designs with the least area overhead when compared to the original designs, the other techniques have similar area overhead, where the minimum (maximum) overhead is computed as $-1.4\%$ ($11.9\%$) on the \textit{b19\_C} (\textit{b20\_C}) instance obtained by the logic locking technique using {\sc xor/xnor} gates (LUTs). 

\begin{table*}[t]
	\centering
	\caption{Synthesis results of designs locked by the state-of-the-art techniques.}
	\vspace{-3mm}
	\footnotesize
	\begin{tabular}{|@{\hskip3pt}l@{\hskip3pt}||c@{\hskip3pt}|@{\hskip3pt}c@{\hskip3pt}|@{\hskip3pt}c@{\hskip3pt}||c@{\hskip3pt}|@{\hskip3pt}c@{\hskip3pt}|@{\hskip3pt}c@{\hskip3pt}||c@{\hskip3pt}|@{\hskip3pt}c@{\hskip3pt}|@{\hskip3pt}c@{\hskip3pt}||c@{\hskip3pt}|@{\hskip3pt}c@{\hskip3pt}|@{\hskip3pt}c@{\hskip3pt}||c@{\hskip3pt}|@{\hskip3pt}c@{\hskip3pt}|@{\hskip3pt}c@{\hskip3pt}|}
		\hline
		\multirow{3}{*}{Instance} & \multicolumn{3}{c||}{ \multirow{2}{*}{Original Circuits} } & \multicolumn{12}{c|}{Locked Circuits} \\
		\cline{5-16}
		& \multicolumn{1}{c}{} & \multicolumn{1}{c}{} & \multicolumn{1}{c||}{} & \multicolumn{3}{c||}{{\sc xor/xnor}} & \multicolumn{3}{c||}{LUT} & \multicolumn{3}{c||}{Anti-SAT} & \multicolumn{3}{c|}{TTLock} \\ 
		\cline{2-16}
		& area & delay & power & area & delay & power & area & delay & power & area & delay & power & area & delay & power \\ 
		\hline \hline
		b17\_C & 23770  & 6468 & 258  & 24035  & 6825 & 273  & 24025  & 6287 & 269  & 24522  & 6355 & 266  & 24715  & 6247 & 269  \\
		b19\_C & 135983 & 6138 & 2071 & 134112 & 6183 & 2052 & 137292 & 6100 & 2132 & 137473 & 6106 & 2088 & 137008 & 6093 & 2087 \\
		b20\_C & 9426   & 5705 & 225  & 9750   & 5823 & 242  & 10554  & 5153 & 265  & 9933   & 5536 & 232  & 10128  & 5365 & 233  \\
		b22\_C & 14166  & 5925 & 327  & 14579  & 5596 & 349  & 15513  & 5982 & 359  & 14558  & 5819 & 330  & 14704  & 5694 & 333  \\
		\hline
	\end{tabular}
	\label{tab:sota_ll}
	\vspace{-4mm}
\end{table*}

\begin{table*}[t]
	\centering
	\caption{Synthesis results of designs locked by ciphers.}
	\vspace{-3mm}
	\footnotesize
	\begin{tabular}{|@{\hskip3pt}l@{\hskip3pt}||l@{\hskip3pt}||c@{\hskip3pt}c@{\hskip3pt}c@{\hskip3pt}||c@{\hskip3pt}c@{\hskip3pt}c@{\hskip3pt}||c@{\hskip3pt}c@{\hskip3pt}c@{\hskip3pt}||c@{\hskip3pt}c@{\hskip3pt}c@{\hskip3pt}|}
		\hline
		\multirow{2}{*}{Technique} & \multirow{2}{*}{Cipher} & \multicolumn{3}{c||}{b17\_C} & \multicolumn{3}{c||}{b19\_C} & \multicolumn{3}{c||}{b20\_C} & \multicolumn{3}{c|}{b22\_C} \\ 
		\cline{3-14}
		& & area & delay & power & area & delay & power & area & delay & power & area & delay & power \\
		\hline \hline
		\multirow{5}{*}{Proposed}  & SIMON\_32\_64\_32                                                     & 32919 & 9735  & 2572  & 138218 & 9959  & 4177  & 19431 & 9861  & 2561  & 24139 & 9649  & 2591 \\
		& SIMON\_64\_96\_42                                                     & 45522 & 11096 & 7354  & 150957 & 11176 & 9157  & 32186 & 10763 & 7382  & 36878 & 10971 & 7580 \\
		& PRESENT\_64\_80\_31                                                   & 48674 & 12578 & 6566  & 162709 & 13751 & 9432  & 42669 & 14247 & 8405  & 40213 & 12487 & 6801 \\
		& PRESENT\_64\_128\_31                                                  & 49852 & 12183 & 9515  & 156288 & 12662 & 11703 & 36354 & 12119 & 9770  & 41841 & 12306 & 9833 \\
		& ASCON\_128\_128\_12                                                   & 75954 & 8539  & 23769 & 180554 & 8669  & 24602 & 62598 & 8808  & 23787 & 67268 & 8570  & 23367 \\
		\hline
		\cite{yasin16_tcad}    & AES                                            & 220968  & 23646  & 31317  & 349141  & 27816  & 32574  & 226357  & 30786  & 30058  & 231139  & 24690  & 30710 \\
		\hline
		\cite{massad22}        & Trivium                                        & 1210130 & 117443 & 811613 & 1287529 & 119188 & 851461 & 1162941 & 119707 & 780385 & 1220987 & 116733 & 822410 \\
		\hline
	\end{tabular}
	\label{tab:lllwc}
	\vspace{-6mm}
\end{table*}

These original circuits are also locked by SOHNI and those using AES~\cite{yasin16_tcad} and Trivium~\cite{massad22}. Table~\ref{tab:lllwc} presents the \mbox{gate-level} synthesis results of locked designs, where \textit{A\_bs\_kl\_r} denotes the LWC algorithm $A$ with its parameters.

Observe from Table~\ref{tab:lllwc} that the use of LWC block ciphers leads to locked designs with significantly less hardware complexity when compared to designs locked by the technique of~\cite{yasin16_tcad} using AES and the technique of~\cite{massad22} using Trivium. For example, the \textit{b20\_C} circuit locked by the techniques of~\cite{yasin16_tcad} and~\cite{massad22} has $3.6\times$ ($6.2\times$) and $18.5\times$ ($31.9\times$) larger area than the same circuit locked by \textit{ASCON\_128\_128\_12} (\textit{PRESENT\_64\_128\_31}), respectively. Furthermore, the same instance locked by the techniques of~\cite{yasin16_tcad} and~\cite{massad22} has $11.6\times$ and $59.8\times$ larger area than the same circuit locked by \textit{SIMON\_32\_64\_32}, respectively. The area overhead on the designs locked by \textit{SIMON\_32\_64\_32} ranges between $106.1\%$, obtained on the \textit{b20\_C} instance, and $1.6\%$, obtained on the \textit{b19\_C} instance, with respect to the original design. 

\begin{figure}[t]
	\vspace*{-6mm}
	\centerline{\includegraphics[width=9.0cm]{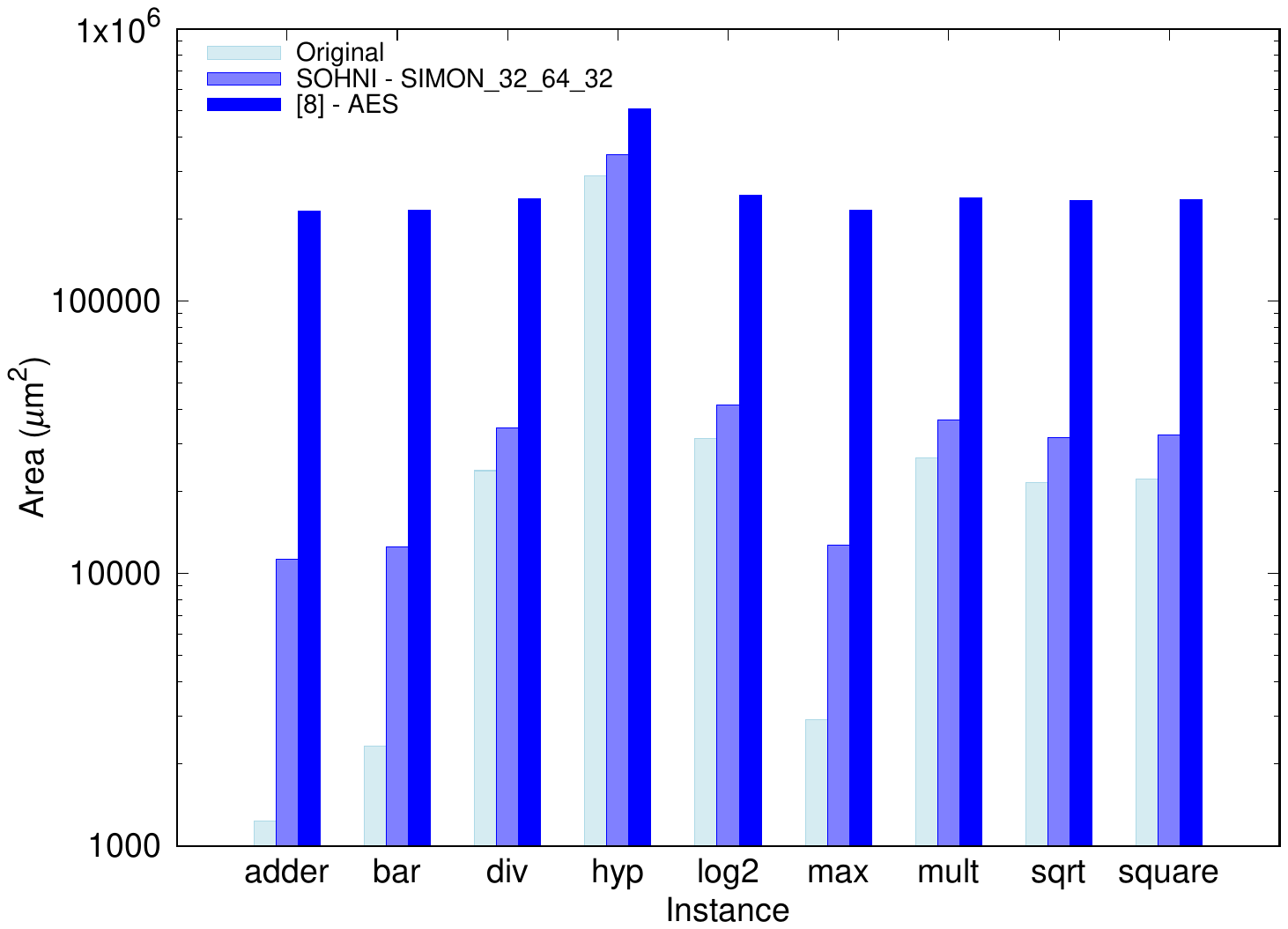}}
	\vspace*{-10mm}
	\caption{Gate-level area of original and locked EPFL circuits.}
	\label{fig:epfl_area}
	\vspace*{-6mm}
\end{figure}

As the second experiment set, the arithmetic circuits of the EPFL benchmark~\cite{amaru15} are used to explore the impact of the proposed techniques on the design area. These circuits are locked by SOHNI using \textit{SIMON\_32\_64\_32} and by the technique of~\cite{yasin16_tcad} using AES. Fig.~\ref{fig:epfl_area} shows the gate-level area values of the original and locked designs. 

Observe that the designs locked by the technique of~\cite{yasin16_tcad} using AES have significantly larger area than those locked by SOHNI using \textit{SIMON\_32\_64\_32} due to the hardware complexity of these block ciphers given in Table~\ref{tab:lwc}. Note that the area overhead in designs locked by SOHNI with respect to the original design reaches up to $9.1\times$ on the \textit{adder} instance that has a small area. This value decreases to $1.4\times$ on the \textit{mult} instance that has a moderate area. On the \textit{hyp} instances that have a large area, this value reduces to $1.2\times$. This experiment indicates that as the complexity of the original design increases, the area overhead in the locked designs decreases, highlighting the scalability of the proposed technique in large size original circuits.

\subsection{Attacks}

Table~\ref{tab:sota_att} presents the run-time of attacks in seconds on the designs locked by the state-of-the-art logic locking techniques given in Table~\ref{tab:sota_ll}. While the SAT-based attack of~\cite{subramanyan15} is run on those locked using {\sc xor/xnor} gates and LUTs, the structural analysis attack KRATT~\cite{aksoy24} is applied to those locked by \mbox{Anti-SAT} and TTLock. The attacks were run with a time limit of 3 days on a computing server, including 32 Intel Xeon processing units at 3.9~GHz and 128~GB memory.

Observe that the existing attacks use a little computational effort to break these state-of-the-art logic locking techniques. This experiment indicates that the techniques that do not rely on provably secure algorithms, such as cryptographic primitives, are broken by existing attacks or will be broken by possible attacks. 

Table~\ref{tab:att} presents the solutions of attacks on designs locked by SOHNI and the previously proposed techniques~\cite{yasin16_tcad, massad22}, where \textit{nok} is the number of key inputs, \textit{REM} and \textit{ALG} are the proposed OG removal and algebraic attacks, respectively, \textit{SAT} includes the OG \mbox{SAT-based} attack~\cite{subramanyan15}, AppSAT~\cite{shamsi17}, and DoubleDIP~\cite{shen17}, and \textit{SAA} includes the structural analysis attacks of~\cite{aksoy24, limaye22}. While the removal attack runs the structural analysis attack of~\cite{aksoy24} on the designs locked by the proposed technique using TTLock, it applies the SAT-based attack~\cite{subramanyan15} to the designs locked by the technique of~\cite{yasin16_tcad} using {\sc xor/xnor} gates. In this table, \textit{NoS} denotes that no solution can be found within a time limit of 3 days while \textit{NA} indicates that the attack is not applicable to the locked design. The solutions of the removal attack are given in seconds.

\begin{table}[t]
	\centering
	\caption{Attacks on designs locked by the state-of-the-art techniques.}
	\vspace{-3mm}
	\footnotesize
	\begin{tabular}{|l||c|c|c|c|}
		\hline
		\multirow{2}{*}{Instance} & {\sc xor/xnor} & LUT       & Anti-SAT & TTLock \\
		\cline{2-5}
		& SAT-based      & SAT-based & KRATT    & KRATT \\ 
		\hline \hline
		b17\_C & 63   & 40   & 59  & 398  \\
		b19\_C & 3270 & 5034 & 414 & 21845\\
		b20\_C & 19   & 23   & 23  & 147  \\
		b22\_C & 47   & 48   & 39  & 219  \\
		\hline
	\end{tabular}
	\label{tab:sota_att}
	\vspace{-6mm}
\end{table}

Observe from Table~\ref{tab:att} that the removal attack can extract the original circuit from the design locked by the technique of~\cite{yasin16_tcad}. Note that it takes the longest time on the \textit{b19\_C} instance since this instance has the largest area. However, the other attacks cannot find the secret key within the given time limit. Hence, this experiment clearly shows that the technique of~\cite{yasin16_tcad} can prevent overproduction, but not piracy. Although the proposed attacks have not been developed for the technique of~\cite{massad22}, the SAT-based attack and its variants and structural analysis attacks cannot find the secret key. On the designs locked by SOHNI, the removal attack cannot identify the original design, and the algebraic attack is not applicable since the output values of the block cipher cannot be found. The \mbox{SAT-based} attack and its variants and structural analysis attacks cannot find the secret key of designs locked by SOHNI. Note that all these attacks are rendered inefficient in all designs due to the compound structure of the proposed technique, indicating that it can prevent both overproduction and piracy. Observe that the proposed technique uses a larger number of key inputs than the others due to the LUT-based obfuscation technique. Finally, we note that similar results were also observed on the locked EPFL circuits.

\begin{table*}[t]
	\centering
	\caption{Attacks on designs locked by ciphers.}
	\vspace{-3mm}
	\footnotesize
	\begin{tabular}{|@{\hskip1.5pt}l@{\hskip1.5pt}||@{\hskip1.5pt}l@{\hskip1.5pt}||c@{\hskip1.5pt}c@{\hskip1.5pt}c@{\hskip1.5pt}c@{\hskip1.5pt}c@{\hskip1.5pt}||c@{\hskip1.5pt}c@{\hskip1.5pt}c@{\hskip1.5pt}c@{\hskip1.5pt}c@{\hskip1.5pt}||c@{\hskip1.5pt}c@{\hskip1.5pt}c@{\hskip1.5pt}c@{\hskip1.5pt}c@{\hskip1.5pt}||c@{\hskip1.5pt}c@{\hskip1.5pt}c@{\hskip1.5pt}c@{\hskip1.5pt}c@{\hskip1.5pt}|}
		\hline
		\multirow{2}{*}{Technique} & \multirow{2}{*}{Cipher} & \multicolumn{5}{c||}{b17\_C} & \multicolumn{5}{c||}{b19\_C} & \multicolumn{5}{c||}{b20\_C} & \multicolumn{5}{c|}{b22\_C} \\ 
		\cline{3-22}
		& & nok & REM & ALG & SAT & SAA & nok & REM & ALG & SAT & SAA & nok & REM & ALG & SAT & SAA & nok & REM & ALG & SAT & SAA \\
		\hline \hline
		\multirow{5}{*}{Proposed} & SIMON\_32\_64\_32       & 448  & NoS          & NA          & NoS   & NoS        & 448  & NoS          & NA          & NoS   & NoS        & 448  & NoS         & NA           & NoS    & NoS       & 448  & NoS          & NA    & NoS        & NoS \\
		& SIMON\_64\_96\_42                                 & 864  & NoS          & NA          & NoS   & NoS        & 864  & NoS          & NA          & NoS   & NoS        & 864  & NoS         & NA           & NoS    & NoS       & 864  & NoS          & NA    & NoS        & NoS \\
		& PRESENT\_64\_80\_31                               & 1029 & NoS          & NA          & NoS   & NoS        & 848  & NoS          & NA          & NoS   & NoS        & 1270 & NoS         & NA           & NoS    & NoS       & 1228 & NoS          & NA    & NoS        & NoS \\
		& PRESENT\_64\_128\_31                              & 1157 & NoS          & NA          & NoS   & NoS        & 1235 & NoS          & NA          & NoS   & NoS        & 1240 & NoS         & NA           & NoS    & NoS       & 1244 & NoS          & NA    & NoS        & NoS \\
		& ASCON\_128\_128\_12                               & 1760 & NoS          & NA          & NoS   & NoS        & 1832 & NoS          & NA          & NoS   & NoS        & 1796 & NoS         & NA           & NoS    & NoS       & 1772 & NoS          & NA    & NoS        & NoS \\
		\hline                                                                                                                                                                                                                                                             
		\cite{yasin16_tcad}       & AES                     & 128  & 14737        & NoS         & NoS   & NoS        & 128  & 40965        & NoS         & NoS   & NoS        & 128  & 12246       & NoS          & NoS    & NoS       & 128  & 16274        & NoS   & NoS        & NoS \\
		\hline                                                                                                                                                                                                                                                             
		\cite{massad22}           & Trivium                 & 80   & NA           & NA          & NoS   & NoS        & 80   & NA           & NA          & NoS   & NoS        & 80   & NA          & NA           & NoS    & NoS       & 80   & NA           & NA    & NoS        & NoS \\
		\hline
	\end{tabular}
	\label{tab:att}
	\vspace{-6mm}
\end{table*}

	\section{Conclusions}
\label{sec:conclusions}

This paper proposed a compound IC protection mechanism that includes a logic locking technique, an LWC block cipher, and a LUT-based obfuscation technique, and presented the developed CAD tool that automates the generation of secure ICs. It also introduced a removal attack that can extract the original circuit from the design locked by a previously proposed technique and an algebraic attack that aims to find the secret key. It was shown that the proposed technique leads to locked designs with significantly less hardware complexity than those generated by previously proposed techniques using cryptography algorithms and mitigates both overproduction and piracy, rendering existing attacks unsuccessful. 
	
	\section*{Acknowledgment}

We thank Mohamed El Massad for providing the codes that generate the restore and perturb units of the DFLT technique with Trivium.

	\bibliography{isqed26}
	\bibliographystyle{IEEEtran}
\end{document}